\documentclass{elsart}
\usepackage{graphicx,amssymb}
\usepackage[authoryear]{natbib}
\journal{New Astronomy}
\begin{document}
\begin{frontmatter}
\title{Constraints on time variation of fine structure constant from WMAP-3yr data}

\author{P. Stefanescu}
\address{Institute for Space Science, Atomistilor 409, Magurele, Ilfov, Romania}
\ead{pstep@venus.nipne.ro}

\begin{abstract}
This paper presents the constraints on the time variation of the fine structure constant at recombination relative to its present value, $\Delta_{\alpha}=(\alpha_{rec}-\alpha_0)/\alpha_0$, obtained from the analysis of the WMAP-3yr Cosmic Microwave Background (CMB) data, with an additional prior on the Hubble 
expansion rate from HST Hubble Key Project. I found out that $-0.039 < \Delta_{\alpha} < 0.010$ at 95\% C.L., which brings a $30\%$ improvement to the previous limits from WMAP-1yr data. The corresponding recombination redshift, $z_{rec}=1\,075_{-63}^{+40}$, shows a delayed recombination epoch compared with the results from WMAP-1yr data.\\
\end{abstract}
\end{frontmatter}

\section{INTRODUCTION}

With the release of WMAP-3yr temperature \citep{Hinshaw06} and polarization data \citep{Page06}, the standard $\Lambda$CDM cosmological model 
has passed another rigorous set of tests, being validated once again 
\citep{Spergel06}. In this era of precision cosmology, the early Universe is 
playing an increasingly more important role as a laboratory where physics not 
accessible on Earth can be tested and verified. Thus, cosmology became an 
excellent tool to answer to many questions of fundamental physics, among which
the time variation of fundamental constants of nature, predicted as 
observational consequences of some currently preferred unification theories 
\citep{Uzan02,Martins02}.

One of the most exciting case is the fine structure constant $\alpha$, possible time variations being already reported from geophysical and astronomical 
observations.

From the Oklo natural nuclear reactor in Gabon, the relative time variation 
of $\alpha$ has been obtained as $-0.9 \times 10^{-7} < \Delta_{\alpha} < 1.2 \times 10^{-7}$ at $z\sim 0.1$ \citep{Damour}.

Terrestrial laboratory measurements give the constraint 
$\Delta \alpha/\alpha<1.4\times10^{-14}$ based on comparisons of rates between 
clocks with different atomic numbers during 140 days of observations 
\citep{Prestage}.

While the results presented above are only upper limits of $\Delta_{\alpha}$, the estimations of its actual value have been firstly computed from spectral analysis of high-redshift quasar absorption systems. Some of the reported values are $\Delta_{\alpha}=(-4.6\pm5.7)\times10^{-5}$ for a 
redshift range $z\sim2-4$ \citep{Varshalovich}, 
$\Delta_{\alpha}=(-1.09\pm0.36)\times10^{-5}$ \citep{Webb00}, these 
determinations being a confirmation of the expectation that the fine structure 
constant is a nondecreasing function of time. From this point of view, the 
result $(\alpha(z_2)-\alpha(z_1))/\alpha_0=(5.43\pm2.52)$ ppm for $z_1=1.15$ and 
$z_2=1.84$ \citep{Levshakov06}, is controversial. In fact, there are also many other contradictory results which, perhaps, may be validated if the fine structure constant would not only be time dependent but also had a spatial variation.  

The CMB presents the advantage of probing directly 
the decoupling epoch ($z \sim 1\,100$) when the $\alpha$ variations from its 
current value are expected to be more important and therefore, more easy to test. A value of $\alpha$ different from $\alpha_0$ at the epoch of recombination changes the ionization evolution at that epoch, inducing effects on the CMB anisotropies observed today.

In the last years several groups have analyzed the variation of fine structure
 constant at recombination using CMB measurements. A lower value of 
$\alpha$ at recombination epoch compared with its present value, was taken into account as a possibility to solve the disagreement between the flat cosmological model and 
the BBN predictions on the one side and the first Doppler peak position and the secondary peaks amplitudes in CMB power spectra obtained by BOOMERANG and MAXIMA on the other side. Negative values of few percent for $\Delta_{\alpha}$ have been reported from BOOMERANG and MAXIMA data analysis \citep{Battye,Avelino00} and from BOOMERANG, DASI and COBE data analysis \citep{Avelino01}.

From the analysis of the first year WMAP data, a $-0.06<\Delta_{\alpha}<0.01$ interval \citep{Rocha031,Rocha032} was reported. Using the same data togheter with HST Hubble Key Project, but considering simultaneous time variations of both $\alpha$ and the electron mass $m_e$, and using a different analysis method, has been obtained the interval $-0.048<\Delta_{\alpha}<0.032$ when only $\alpha$ varies \citep{Ichikawa06}.

In the present work, the WMAP-3yr CMB data are being analyzed in order to find new 
limits on the $\alpha$ value at recombination, and the corresponding limits on 
the recombination redshift.\footnote{Throughout this paper, we consider the recombination redshift as being the redshift corresponding to maximum of the visibility function.}

In section 2 it is reviewed the standard recombination process and are described the changes involved by a different value of the fine structure constant at recombination.

Section 3 presents the results of the analysis of WMAP-3yr data in the context of the recombination process affected by time varying fine structure constant.
 
In section 4 are summarized the conclusions of the work.

\section{TIME VARYING {\Large{$\alpha$}} AND CMB}

It is well known the effect of changing the value of the fine structure 
constant on the energy levels of Hydrogen atom and, therefore, on the photon frequencies corresponding to the transitions between two such levels. As a consequence, a change in $\alpha$ 
value will change the dynamics of the Hydrogen recombination process in the 
Universe, which affects the CMB fluctuations observed today. As the CMB primary anisotropies probe the recombination epoch, it is natural to 
consider imposing limits on possible variation in $\alpha$ value at the recombination 
epoch relative to present value using CMB anisotropy data.

In the following I briefly review the Hydrogen and Helium recombination process and the implications of non-standard value of $\alpha$, neglecting the impact on Helium.

Before recombination, the photon gas was coupled to electron-baryon fluid through Thomson scattering on free electrons, the cross section for this process being given by \citep{Weinberg}

\begin{equation}
\sigma_T=\frac{1}{6\pi}\frac{e^4}{m_e^2}\propto\alpha^2.
\end{equation}

The CMB formation followed the recombination process, when the photons became free particles after electrons have been captured by ions.

For modeling the recombination process I used the treatment implemented in the RECFAST code \citep{Seager}. The equations describing the evolution of proton fraction $x_p$, the singly ionized Helium fraction $x_{HeII}$ and the matter temperature $T_M$ are:

\setlength\arraycolsep{1pt}
\begin{eqnarray}
\frac{dx_p}{dz}&=&\nonumber\\
& &\frac{C_H}{H(z)(1+z)}[x_ex_pn_HR_H-\beta_H(1-x_p)e^{-h\nu_H/kT_M}],\\
\frac{dx_{HeII}}{dz}&=&\frac{C_{He}}{H(z)(1+z)}\nonumber\\
& &\times[x_{HeII}x_en_HR_{HeI}-\beta_{HeI}(f_{He}-x_{HeII})e^{-h\nu_{HeI}/kT_M}],\\
\frac{dT_M}{dz}&=&\nonumber\\
& &\frac{8\sigma_Ta_RT_R^4}{3H(z)(1+z)m_e}\frac{x_e}{1+f_{He}+x_e}(T_M-T_R)+\frac{2T_M}{(1+z)}.
\end{eqnarray}

In the above equations, $H(z)$ is the Hubble expansion rate at redshift $z$, $h$ is the Planck constant, $k$ is the Boltzmann constant, $c$ is the speed of light, $a_R=k^4/(120\pi c^3h^3)$ is the blackbody constant, $n_H$ is the Hydrogen number density, $x_p=n_e/n_H$ is the proton fraction, $x_{HeII}=n_{HeII}/n_H$ is the singly ionized Helium fraction and $x_e=n_e/n_H=x_p+x_{HeII}$ is the electron fraction. The number fraction of Helium to Hydrogen is $f_{He}=Y_p/(4(1-Y_p))$, where $Y_p=0.24$ is the primordial Helium mass fraction. The radiation temperature $T_R(z)=T_{CMB}(1+z)$ is identical to $T_M$ at high redshift because of the coupling of photons and baryons through Thomson scattering. $R_H$ is the case B recombination coefficient for H, and is given by the fit formula

\begin{equation}
R_H=F10^{-19}\frac{at^b}{1+ct^d}m^3s^{-1}
\end{equation}

where $t=T_M/(10^4)K$, $a=4.309$, $b=-0.6166$, $c=0.6703$, $d=0.5300$ \citep{Pequignot} and $F=1.14$ is the fudge factor \citep{Seager} introduced in order to reproduce the results of the multilevel calculation by speeding up recombination in the standard scenario. $\beta_H$ is the photoionization coefficient

\begin{equation}
\beta_H=R_H\left(\frac{2\pi m_ekT_M}{h^2}\right)^{\frac{3}{2}}exp(-\frac{B_{H2s}}{kT_M}),
\end{equation}

and $C_H$ is the Peebles reduction factor

\begin{equation}
C_H=\frac{[1+K_H\Lambda_Hn_H(1-x_p)]}{[1+K_H(\Lambda_H+\beta_H)n_H(1-x_p)]},
\end{equation}

which accounts for the presence of non-thermal Ly-$\alpha$ resonance photons. In the above, $B_{H2s}=3.4$eV is the binding energy in the $2s$ energy level, $\nu_H=(B_{H1s}-B_{H2s})/h$ is the Ly-$\alpha$ frequency, $\Lambda_H$ is the rate of decay of the $2s$ excited state to the ground state via $2$ photons, and $K_H=c^3/(8\pi \nu_H H(z))$.

The quantities $R_{HeI}$, $\beta_{HeI}$, $C_{He}$ and $\nu_{HeI}$ from Eq. (3) are the analogous for Helium of the quantities from Eq. (2) and their expressions may be found, for example, in Ichikawa et al. (2006). In this work, the small effect of changing $\alpha$ on Helium recombination process has been neglected.

In order to take into account the changing value of $\alpha$ in the recombination process, I have modified the evolution equations for proton fraction and for matter temperature, considering their $\alpha$ dependence. The quantities which depends on $\alpha$ in these equations have been Taylor expanded up to first order in $\Delta_{\alpha}$ according to their scaling relations \citep{Kapling98}
\setlength\arraycolsep{1pt}
\begin{eqnarray}
R_H&\propto&\alpha^{2(1+\xi)}, \nonumber\\
B_{Hn}&\propto&\alpha^2, \nonumber \\
K_H&\propto&\alpha^{-6}, \nonumber \\
\Lambda_H&\propto&\alpha^8,
\end{eqnarray}
where $\xi=0.7$ was adopted. 

\section{CMB CONSTRAINTS ON {\Large{$\alpha$}} VARIATION USING LATEST WMAP DATA}

In this work, in order to search for new CMB limits on $\alpha$ value 
at recombination, I have analyzed the WMAP-3yr CMB anisotropy data 
\citep{Hinshaw06,Page06} in the framework of the extended cosmological model which 
includes the variation of the fine structure constant at recombination with 
respect to its present value. The data analysis has been done using Markov Chain Monte Carlo (MCMC) techniques \citep{Mackay} implemented in the COSMOMC code \citep{Lewis02}.

For this purpose I modified the RECFAST code to compute the Hydrogen 
recombination in the hypothesis of different value of $\alpha$ at last 
scattering. The relative variation $\Delta_{\alpha}=(\alpha_{rec}-\alpha_{0})/\alpha_0$ was added as an additional parameter; the modified equations for the evolution of Hydrogen ionization fraction and matter temperature have been integrated with CAMB code \citep{Lewis00}, used by COSMOMC to compute the theoretical CMB power spectra.

Together with $\Delta_{\alpha}$ the following cosmological parameters have been varied: physical density in baryons $\Omega_bh^2$, physical density in DM 
$\Omega_{DM}h^2$, Hubble constant $H_0$, reionization redshift $z_{re}$, 
spectral index $n_s$ and amplitude $\Delta_R^2$ of primordial fluctuations; the cosmological constant $\Omega_{\Lambda}$ and the optical depth to reionization $\tau$ have been derived. To realistically constrain 
the Hubble expansion rate taking into account its degeneracy with 
$\Delta_{\alpha}$ \citep{Hannestad}, the HST Key Project prior \citep{Freedman01,Lewis02,Rubino03} has been used in addition to WMAP-3yr data.

The modified version of COSMOMC software package has been run on 8 Markov chains, using the "variance of chain means"/"mean of chain variances" R statistic \citep{Brooks98} as convergence criterion with the choise $R-1<0.03$. 

The most likely values of cosmological parameters obtained from MCMC 
simulations are given in Table 1. The mean values of the 
standard cosmological parameters are in the limits reported
 by WMAP team \citep{Spergel06}.

\begin{center}
Table 1: $\Lambda CDM$ with varying $\alpha$ model parameters and their 68\% confidence intervals obtained from WMAP-3yr data.

\small
\begin{tabular}{ccc}
\hline   \hline
Parameter & Mean & ML \\ \hline
$100\Omega_bh^2$ & $2.15_{-0.11}^{+0.11}$ & 2.18 \\
$\Omega_{DM}h^2$ & $0.104 _{-0.008}^{+0.008}$ & 0.108 \\
$H_0$ & $68.43_{-6.9}^{+6.5}$ & 68.74 \\
$z_re$ & $11.37_{-2.56}^{+2.58}$ & 12.02 \\
$n_s$ & $0.96_{-0.02}^{+0.02}$ & 0.96 \\
$10^{10}\Delta_R^2(k=0.05)$ & $20.93_{-1.44}^{+1.44}$ & 21.39 \\
Age/GYr & $14.24_{-0.66}^{+0.72}$ & 14.05 \\
$\Delta_{\alpha}$ & $-0.011_{-0.006}^{+0.017}$ & -0.006 \\
$\Omega_{\Lambda}$ & $0.72_{-0.05}^{+0.06}$ & 0.73 \\
$\Omega_m$ & $0.27_{-0.05}^{+0.06}$ & 0.27 \\
$\tau$ & $0.090_{-0.02}^{+0.03}$ & 0.096 \\ \hline 

\end{tabular}
\normalsize
\end{center}

The marginalized distributions of the parameters obtained from simulations are 
presented in Figure 1. The observed differences between the curves corresponding to the parameters $H_0$, $\Omega_bh^2$, $\Omega_m$, $\Omega_{\Lambda}$ and the age of the Universe 

\begin{center}
\begin{figure}
\includegraphics[height=8.cm]{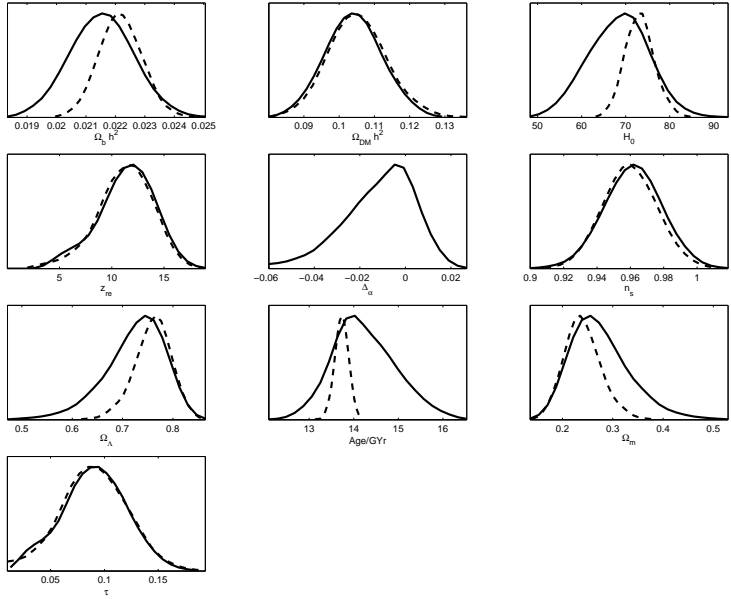}
\caption{Marginalized distributions for the parameters of $\Lambda$CDM model with varying $\alpha$ obtained from the analysis of WMAP-3yr data and HST Key 
Project (continuous lines) compared with the distributions of the parameters in the standard $\Lambda$CDM model (dashed lines) from the same data.}
\end{figure}
\end{center}

in the $\Lambda$CDM with varying $\alpha$ model compared to those of the standard $\Lambda$CDM model  are due to the degeneracy of these parameters with $\alpha$.

Figure 2 presents the 2D marginalized distributions for pairs of parameters at 
68\% and 95\% CL for $\Lambda$CDM with varying $\alpha$ and the  $\Lambda$CDM models. The negative values for $\Delta_{\alpha}$ in the $\Lambda$CDM with varying $\alpha$ model are correlated with smaller values of $H_0$ and $\Omega_bh^2$ then in the $\Lambda$CDM model. On the other side, the distributions for $\Omega_{DM}h^2$ are almost the same in both cases, therefore, in the case with varying $\alpha$ the parameter $\Omega_m=\Omega_b+\Omega_{DM}$ tends to go to higher values and $\Omega_{\Lambda}$ to lower values then in the case with no $\alpha$ variations. Consequently, the distribution for the age of the Universe is widened, with mean value larger then in the case with constant $\alpha$.

Figure 3 shows the $C_l^{TT}$, $C_l^{EE}$ and $C_l^{TE}$ CMB power spectra for the best fit parameters of the model with varying $\alpha$ compared with the best fit CMB power spectra for the $\Lambda$CDM model and the WMAP-3yr experimental data. It is remarkable that the two best fit models are almost identical from the CMB power spectra point of view.

The 95 \% confidence interval for the additional parameter obtained from 
WMAP-3yr data is

\begin{center}
\begin{figure}
\includegraphics[height=14.cm]{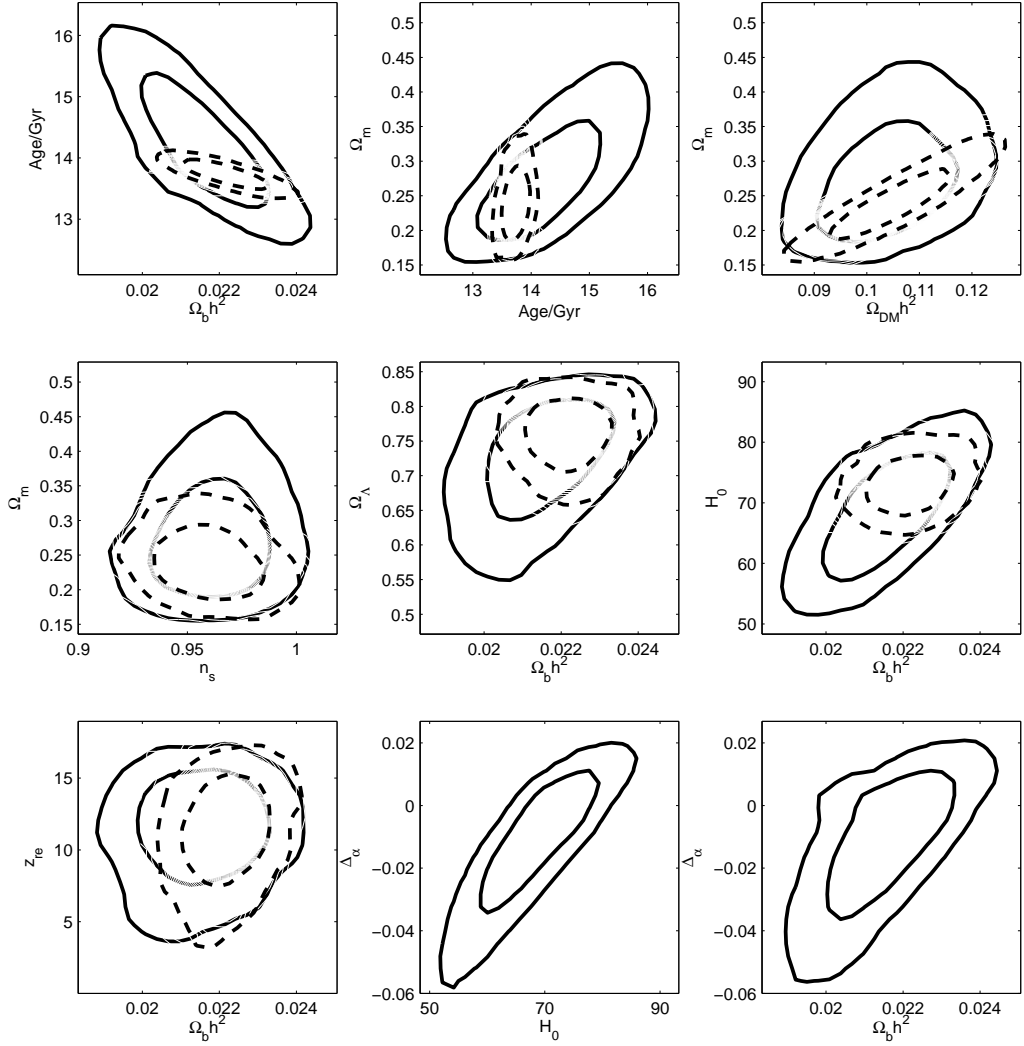}
\caption{2D - marginalized distributions for the parameters of  $\Lambda$CDM model with varying $\alpha$ obtained from the analysis of WMAP-3yr data and HST Key Project (continuous lines) compared with the 2D - distributions of the parameters in the standard $\Lambda$CDM model from the same data (dashed lines). The contours are at 68\% and 95\% CL.}
\end{figure}
\end{center}

\begin{eqnarray}
\\
-0.039 < \Delta_{\alpha} < 0.010,\nonumber
\end{eqnarray}
 which is a tighter CMB constraint on the variation of fine structure constant 
at recombination than previously reported from the analysis of WMAP-1yr data \citep{Rocha031,Rocha032,Ichikawa06}.

The interval for the recombination redshift corresponding to the limits on $\Delta_{\alpha}$,
\begin{eqnarray}
1\,012 < z_{rec} < 1\,115,\nonumber
\end{eqnarray}

\begin{center}
\begin{figure}
\includegraphics[width=14.cm,height=10.5cm]{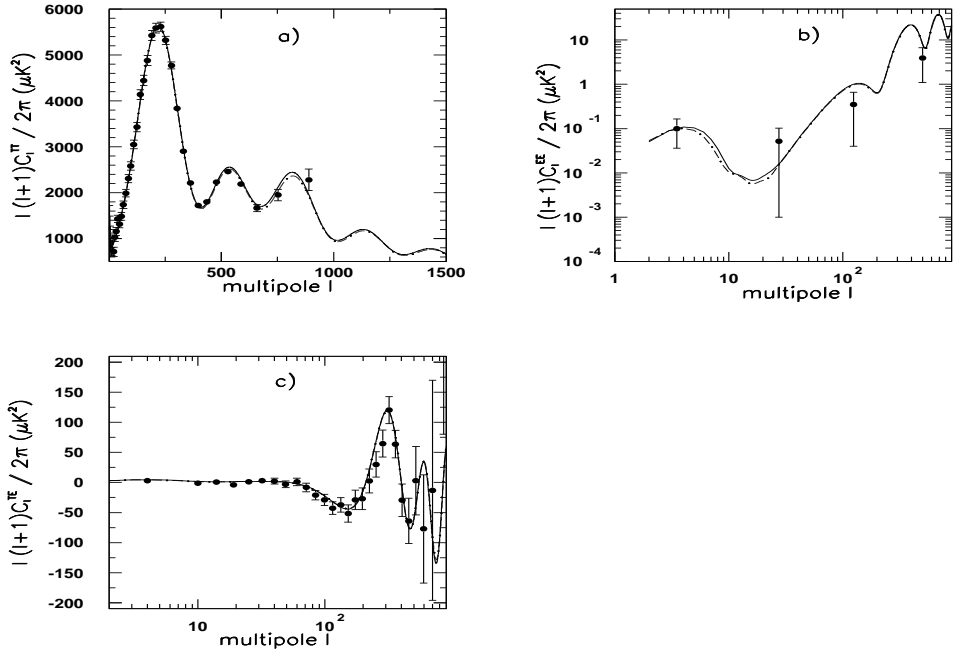}
\caption{CMB power spectra $C_l^{TT}$ (a), $C_l^{EE}$ (b) and $C_l^{TE}$ (c) for the best fit $\Lambda$CMD model with varying $\alpha$ (continous lines) and for the the best fit $\Lambda$CDM model with constant $\alpha$ (computed by the WMAP team \citep{lambda}) (dash-dotted lines) and the WMAP-3yr binned data and their errors.}
\end{figure}
\end{center}

is tighter compared to those corresponding to limits on $\Delta_{\alpha}$ from WMAP-1yr data analysis \citep{Rocha031,Rocha032,Ichikawa06}.

 The recombination redshift corresponding to the best fit parameters of the model with varying $\alpha$, $z_{rec}=1\,075$, indicate a delayed recombination compared with the results from WMAP-1yr data.

 The relative variation of $\alpha$ in redshift unit corresponding to the best 
fit is
\begin{eqnarray}
\alpha^{-1}d\alpha/dz = -5.954 \times 10^{-6}\nonumber
\end{eqnarray}
or, equivalent, in time unit,
\begin{eqnarray}
\alpha^{-1}d\alpha/dt = -4.65 \times 10^{-13} yr^{-1},\nonumber
\end{eqnarray}
both in agreement with the predicted constraints on $\alpha$ from CMB \citep{Hannestad,Kapling98}.

\section{CONCLUSIONS}

Based on the analysis of WMAP-3yr data with MCMC techniques, I obtained a tighter CMB constraint on the value of fine structure constant at recombination epoch then was previously obtained from WMAP-1yr data. The results confirm the limits predicted to be achieved from CMB.

The WMAP-3yr constraints on other cosmological parameters like $H_0$, $\Omega_bh^2$, $\Omega_m$ and $\Omega_{\Lambda}$ are relaxed because of their degeneracy with the $\alpha$ variation, confirming that a non-standard recombination weakens the constraints on other cosmological parameters \citep{Bean07}.

Using the constraint on $\alpha$ variation, I have also obtained the recombination
 redshift, finding a delay of the recombination epoch, similar to that of 
the reionization epoch and of the structure formation computed from WMAP-3yr versus WMAP-1yr data \citep{Popa}.

\section*{Acknowledgmets}

The author thanks L.A. Popa, A. Vasile and O.M. Tantareanu for useful discussions and suggestions. She also acknowledges the Cosmic Rays and Nuclear Astrophysics group of ISS for providing the computing facilities.

\end{document}